\documentclass[runningheads]{llncs}
\usepackage{graphicx}
\usepackage{pdfpages}
\usepackage{booktabs}
\usepackage{cite}

\begin{document}
\title{Segmentation of the Carotid Lumen and Vessel Wall using Deep Learning and Location Priors}
\author{Florian Thamm \inst{1,2} \and Felix Denzinger \inst{1,2} \and Leonhard Rist \inst{1,2} \and Celia Martin Vicario \inst{1,2} \and Florian Kordon \inst{1} \and Andreas Maier \inst{1}}
\authorrunning{F. Thamm et al.}
\institute{Pattern Recognition Lab, Friedrich-Alexander-University Erlangen-Nuremberg, Erlangen, Germany \\  \and
Learning Approaches for Vascular Analysis Group, Friedrich-Alexander-University Erlangen-Nuremberg, Erlangen, Germany}
\maketitle              %
\begin{abstract}
In this report we want to present our method and results for the Carotid Artery Vessel Wall Segmentation Challenge. We propose an image-based pipeline utilizing the U-Net architecture and location priors to solve the segmentation problem at hand. 

\end{abstract}

\section{Introduction}
One of the leading causes of death worldwide is atherosclerosis, a disease which is defined as the build-up of plaque deposits and formations narrowing the lumen, affecting medium and large-sized arteries \cite{herrington2016epidemiology, mitrovska2009atherosclerosis}. This is especially critical for the carotids, which supply the brain and face together with the vertebral arteries. Occlusion in those vessels can thus cause severe irreversible damage if not treated immediately. Therefore, the diagnosis of atherosclerotic vessels is of crucial importance. The vessel wall imaging (VWI) done with magnetic resonance imaging (MRI) is a modality designed to identify and localize stenosis of the atherosclerotic kind. In particular, the fast 3D carotid black blood MRI sequence (3D Motion Sensitized Driven Equilibrium prepared Rapid Gradient Echo, short 3D-MERGE) is designed to visualize blood vessels with sub-millimeter isotropic resolution and at the same time atherosclerotic malformations, their severity, size, and the overall stenosis \cite{balu2011carotid}. Despite the good visibility using the 3D-MERGE sequence, expert knowledge is still required to read such an image. As this is a tedious and time consuming task, the Vessel-Wall-Segmentation Challenge hosted by the Vascular Imaging Laboratory from the University of Washington has been organized to foster and encourage automated methods addressing the segmentation of the outer and the inner (lumen) wall of the left and right external (ECAL, ECAR) and internal carotids (ICAL, ICAR). In this report, we present our approach for the automated vessel wall segmentation and discuss the achieved results.

\section{Methods}
The following section describes the methods we used to solve the segmentation task at hand. First, we propose a processing pipeline utilizing the U-Net \cite{ronneberger2015u} architecture which receives the MRI slices and returns the contours based on the segmentation. Furthermore, we describe all hyperparameters of the network and its training. Secondly, we describe certain characteristics in the application of said pipeline and introduce two concepts our pipeline follows to increase the segmentation accuracy.

\subsection{Pipeline}
\label{sec:pipeline}
\begin{figure}[h]
    \label{fig:pipeline}
  \centering
  \includegraphics[width=\textwidth]{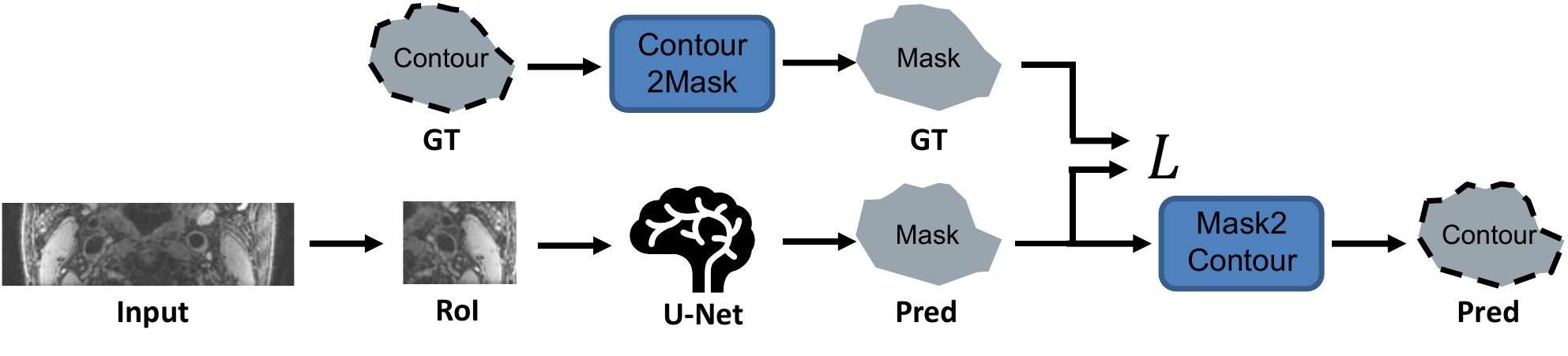}
  \caption{Proposed Pipeline. Training includes all paths leading to the Loss $L$. Inference (testing) only includes the bottom path from the input to the predicted contour.}
\end{figure}

The illustration in Fig.~\ref{fig:pipeline} shows both, training and testing phase of the proposed pipeline. During training, a 2D U-Net model receives a region of interest (RoI) of the input slices and predicts a binary segmentation mask of the carotids, the lumen, and wall in separate channels. The RoI extraction is described in Sec.~\ref{sec:concepts}. The internal and external carotid arteries were treated separately with two different models but the same architecture. All ground truth (GT) masks were given as contour formats in the CASCADE file format \cite{kerwin2007magnetic}, thus it is necessary to convert the contours into a binary mask format, yielding the ground truth mask. This mask is used to supervise the architecture with the binary cross-entropy loss $L$. During the inference and thus testing, the predicted binary mask is transferred back into a contour-based format in a post-processing step. This format is used for the final evaluation. As described later in Sec. \ref{sec:concepts}, the model predicts both sides (left and right) separately. 
The U-Net architecture, consisting of 4 downsampling and hence 4 upsampling steps, beginning with 64 channels and doubling channels up to the deepest stage of 1024 channels, was trained using Adam for 1500 epochs using a learning rate of $10^{-4}$ with a batch size of 32. Max-pooling was used for the downsampling path and transposed Convolutions for the upsampling path.

\subsection{Concepts}
\label{sec:concepts}
\paragraph{Contour-Mask Consistency}
All ground truth masks were given as contours in the CASCADE file format \cite{kerwin2007magnetic}. As described in \ref{sec:pipeline} the pipeline we propose is image-based, which means, the contours are predicted based on a binary segmentation of the internal and external carotid inner and outer walls. As the ground truth masks need to be converted to masks, and the predictions vice-versa into contours, it is hence of importance that both transforming routines are cycle consistent, explained exemplary in the following. Suppose some contour $A$ is given and the task is now to transform that contour into its mask representation using a function \texttt{Contour2Mask}. If this mask is now used to be transformed back into the contour representation using a function \texttt{Mask2Contour} returning contour $B$, the Contour-Mask Consistency states that $B = A$. To enforce this consistency contour points are forced to lie on the image grid, before converting them into binary masks.
\paragraph{Reduction to Region of Interest (RoI)}
To counteract class imbalance, as it is inherent in the problem at hand, we propose to reduce the input image to the region where carotids are expected. This can be done for the left and right carotids respectively. To evaluate this region, the smallest bounding box was determined containing all ground truth segmentation of one side. This bounding box was additionally enlarged to ensure outliers in the test data are covered as well, to a final size of $160 \times 160$ visualized as RoI in \ref{fig:pipeline}. The U-Net segments both bounding boxes, left and right, separately from each other. This has the advantage that either one side can be flipped always to the opposite side making left and right boxes visually look the same, since a symmetrical anatomy can be assumed, or, to augment and flip left and right randomly. Small experiments have shown the latter results in better validation performances.

\section{Evaluation}
The final evaluation was done based on 20 datasets with 720 voxels in height (x), width (y), and depth (z). Five datasets had 640 voxels in all dimensions instead. In total 25 datasets were therefore available. Due to technical issues, the initial submission was not robust against resolutions different to 720$^3$ voxels, thus failed for five datasets in total 107 slices. This has been fixed and the evaluation restarted without retraining, which leads to the performance shown in Tab.~\ref{tab:results}. %
\ifx true false
\begin{table}[]
\label{tab:results}
\centering
\begin{tabular}{llll}
\hline
Metric                                                                                      & \begin{tabular}[c]{@{}l@{}}Initial\\ Submission\end{tabular} & \begin{tabular}[c]{@{}l@{}}Resolution\\ Robust\end{tabular} & \begin{tabular}[c]{@{}l@{}}Dynamic\\ Bounding Boxes\end{tabular} \\ \hline
Dice Score                                                                                  & 0.672±0.248                                                  & 0.707±0.238                                                 & \textbf{0.720±0.220}                                             \\
Lumen area difference                                                                       & 0.142±0.239                                                  & \textbf{0.110±0.208}                                        & 0.119±0.228                                                      \\
Wall area difference                                                                        & 0.162±0.224                                                  & \textbf{0.139±0.205}                                        & 0.144±0.219                                                      \\
Normalized wall index difference                                                            & 0.141±0.212                                                  & 0.135±0.233                                                 & \textbf{0.127±0.215}                                             \\
\begin{tabular}[c]{@{}l@{}}Hausdorff distance on \\ lumen normalized by radius\end{tabular} & 1.095±4.813                                                  & \textbf{1.058±4.705}                                        & 1.060±5.097                                                      \\
\begin{tabular}[c]{@{}l@{}}Hausdorff distance on \\ wall normalized by radius\end{tabular}  & 0.810±2.953                                                  & 0.845±3.007                                                 & \textbf{0.804±2.981}                                             \\
Quantitative score                                                                          & 0.633±0.335                                                  & \textbf{0.691±0.309}                                        & 0.645±0.348                                                      \\
Number of unmatched slices                                                                  & 371                                                          & \textbf{264}                                                & 425                                                              \\ \hline
\end{tabular}
\end{table}
\fi

\begin{table}[]
\label{tab:results}
\centering
\begin{tabular}{lcc}
\hline
Metric                                                                                      & \begin{tabular}[c]{@{}l@{}}Initial\\ Submission\end{tabular} & \begin{tabular}[c]{@{}l@{}}Resolution\\ Robust\end{tabular}\\ \hline
Dice Score                                                                                  & 0.672±0.248                                                  & \textbf{0.707±0.238}                                                                                        \\
Lumen area difference                                                                       & 0.142±0.239                                                  & \textbf{0.110±0.208}                                                                                           \\
Wall area difference                                                                        & 0.162±0.224                                                  & \textbf{0.139±0.205}                                                                                            \\
Normalized wall index difference                                                            & 0.141±0.212                                                  & \textbf{0.135±0.233}                                                                                          \\
\begin{tabular}[c]{@{}l@{}}Hausdorff distance on \\ lumen normalized by radius\end{tabular} & 1.095±4.813                                                  & \textbf{1.058±4.705}                                                                                            \\
\begin{tabular}[c]{@{}l@{}}Hausdorff distance on \\ wall normalized by radius\end{tabular}  & \textbf{0.810±2.953}                                                  &0.845±3.007                                                                                            \\
Quantitative score                                                                          & 0.633±0.335                                                  & \textbf{0.691±0.309}                                                                                            \\
Number of unmatched slices                                                                  & 371                                                          & \textbf{264}                                                                                                            \\ \hline
\end{tabular}
\end{table}

\section{Summary}
For the Carotid Artery Vessel Wall Segmentation Challenge, we have proposed a U-Net-based pipeline that automatically performs a segmentation of the internal and external carotids on both sides based on 3D-MERGE MRI sequences. The network is supervised in the image domain, while the transfer from- and to contours is cycle consistent. To counteract class imbalances in the segmentation, the pipeline automatically, and dynamically extracts bounding boxes, at the regions where the carotids are expected. In the final test evaluation, our approach achieved a dice score of 0.707 and a quantitative measure of 0.691. 

\section*{Author Contribution Statement}
The authors confirm contribution to the work as follows. Conception and management: F.T., F.D., Manuscript draft: F.T., Manuscript review: F.D., L.R., C.V., A.M., Preparation and recording of presentations: F.T., Training of models: F.D., Pipeline implementation: F.D., L.R., F.T., F.K., C.V., Deployment: F.T., L.R., Evaluations: F.T., F.D., Various experiments: F.D., F.K., F.T., L.R., C.V.

\bibliographystyle{splncs04}
\bibliography{bibliography}
\end{document}